\pdfoutput=1
\documentclass[aps,prc,twocolumn,floatfix,superscriptaddress,nofootinbib,preprintnumbers,longbibliography,amsmath,amsthm,amssymb]{revtex4-2}
\usepackage[dvipdfmx]{graphicx}
\usepackage{amsfonts}
\usepackage{color}
\usepackage[colorlinks=true,linkcolor=blue,citecolor=blue]{hyperref}
\usepackage{dcolumn}
\usepackage{bm}
\usepackage{braket}
\usepackage[normalem]{ulem}  

\begin{document}
\allowdisplaybreaks[1]
\title{Correlations of $Q_{\beta}$-values with symmetry energy and effective mass studied within Skyrme energy--density functionals}

\author{Futoshi Minato}
\email[E-mail: ]{minato.futoshi.009@m.kyushu-u.ac.jp}
\affiliation{Department of Physics, Kyushu University, Fukuoka 819-0395, Japan}
\affiliation{RIKEN Nishina Center for Accelerator-Based Science, Wako, Saitama 351-0198, Japan}
\affiliation{Nuclear Data Center, Japan Atomic Energy Agency, Tokai, Ibaraki 319-1195, Japan}

\author{Yifei Niu} 
\email[E-mail: ]{niuyf@lzu.edu.cn}
\affiliation{MOE Frontiers Science Center for Rare Isotopes, Lanzhou University, Lanzhou 730000, China}
\affiliation{School of Nuclear Science and Technology, Lanzhou University, Lanzhou 730000, China}

\author{Kenichi Yoshida}
\email[E-mail: ]{kyoshida@rcnp.osaka-u.ac.jp}
\affiliation{Research Center for Nuclear Physics, the University of Osaka, 
Ibaraki, Osaka 567-0047 Japan}
\affiliation{RIKEN Nishina Center for Accelerator-Based Science, Wako, Saitama 351-0198, Japan}
\affiliation{
 Center for Computational Sciences, University of Tsukuba, Tsukuba, Ibaraki 305-8577, Japan
}

%
\date{\today}
\begin{abstract}

\begin{description}
\item[Background]
The $\beta$-decay half-lives of nuclei are sensitive to the values of $Q_{\beta}$. 
For accurate theoretical predictions, it is essential to develop an effective interaction or an energy density functional (EDF) that can systematically reproduce experimental $Q_{\beta}$ values.
\item[Purpose]
The challenge lies in identifying an appropriate EDF for an accurate $Q_{\beta}$ prediction. 
To address this, we focus on the bulk properties of nuclei that have correlations with $Q_{\beta}$. 
The primary objective of this study is to determine which nuclear bulk properties are sensitive to $Q_{\beta}$, providing information on the key nuclear characteristics that influence $\beta$-decay calculations.
\item[Method]
We employ the Skyrme energy-density functionals to find correlations between $Q_{\beta}$ and the nuclear bulk properties, assuming spherical symmetry.
Using $42$ different Skyrme EDFs, we analyze these correlations by evaluating Pearson linear coefficients, focusing particularly on the relationship between $Q_{\beta}$ and various nuclear properties.
\item[Results]
We found that the symmetry energy at low densities shows a correlation with the $Q_{\beta}$ value. 
In particular, this correlation becomes stronger for functionals with an effective mass close to $1$. 
However, as the nuclear density increases, the correlation weakens.
\item[Conclusions]
From our analysis, we found that a symmetry energy of $32.8\pm0.7$~MeV and effective mass of $m^{*}/m\ge0.75$ at the saturation density is the most likely to systematically reproduce the experimental data of $Q_{\beta}$ systematically.
\end{description}
\end{abstract}
\maketitle
\section{Introduction}
\label{sect:intro}
One of the main challenges in nuclear theoretical study is to systematically derive the ground-state properties from light to heavy nuclei within a certain nuclear model.
Energy-density functional (EDF) models are undoubtedly the most widely applied method to address this problem.
In these models, various effective interactions have been developed for both non-relativistic (Skyrme- and Gogny-type) and relativistic frameworks~\cite{VautherinBrink, Bender2003, Guo2024, Grams2025}.
Recently, Hartree-Fock methods with chiral effective interactions have also been advancing, although their application remains limited to specific nuclei in the nuclear chart~\cite{PhysRevC.94.014303, PhysRevC.101.044309, PhysRevC.107.064307}.
In the early stage of nuclear EDF development, effective interactions were determined primarily to reproduce the ground-state properties of a few selected nuclei and certain nuclear matter properties. However, they are now subject to a broader range of constraints, including neutron-rich nuclei and symmetry energies of nuclear matter~\cite{Nakada2013, Colo2014, Inakura2015, Margueron2018, Roca-Maza2018, Lattimer2023}.
\par
EDF models have enabled us to estimate various nuclear ground-state properties across a wide region of the nuclear chart. Moreover, they allow us to investigate excited states by employing additional models such as the random-phase approximation~\cite{Yoshida:2013bma, Minato2022}, particle-vibrational coupling~\cite{Yoshida2023, Yniu2012}, the generator-coordinate method~\cite{Bender2008}, and others.
The potential of EDF models is particularly beneficial for applications in the study of $r$-process nucleosynthesis~\cite{B2FH,Cameron1957}, which requires comprehensive information on unstable nuclei (for a recent review on the $r$-process, see e.g.~\cite{Kajino2019, RevModPhys.93.015002}).
Key nuclear properties that sensitively affect $r$-process abundance patterns are mass, neutron-capture rates, and $\beta$-decay.
Among these, $\beta$-decay plays a role in determining the time-scale of $r$-process, significantly impacting the resulting abundance pattern.
%
\par
Theoretical studies on $\beta$-decay have been actively carried out by various groups, applying various kinds of effective interactions to $\beta$-decay calculations~\cite{Dongdong2014, Yoshida2015, Niu2015, Wang2016, Sarriguren2016, Tomislav2016, Suzuki2018, Yoshida2019, Ney2020, Borzov2020, Nomura2020, Minato2022}.
Despite these efforts, there is still no clear indication of which type of effective interaction is most appropriate for $\beta$-decay calculations.
This difficulty arises from the fact that $\beta$-decay in most nuclei is triggered by external fields involving spin operators (e.g. the Gamow-Teller transition), which is odd in the time-reversal symmetry.
In general, effective interactions used in non-relativistic density functional models are determined based on the ground-state properties of even-even nuclei that are even in the time-reversal symmetry. 
As a result, nuclear responses to external fields, including spin operators, are not uniquely determined.
However, it is known that one of the Landau parameters, $G_{0}'$, derived from nuclear matter limit of effective interactions, play a key role in producing Gamow-Teller states at the correct energies and strengths~\cite{Giai1981}.
Similarly, it was pointed out that the amount of low-lying Gamow-Teller strength also depend on the value of $G_{0}'$~\cite{Bender2002}.
These findings suggest that $G_{0}'$ is correlated with $\beta$-decays half-lives.
Indeed, the SLy4~\cite{SLy}($G_{0}'=0.881$) and the SkO'~\cite{SkO}($G_{0}'=0.792)$ forces have been shown to reproduce reasonable $\beta$-decay half-lives, whereas the SLy5 force~\cite{SLy}, which has a smaller $G_{0}'$($=-0.141$), severely underestimates half-lives.
\par
According to the Fermi's golden rule for $\beta$-decays, the half-life is proportional to the fifth power of the $\beta$-decay $Q$-value ($Q_{\beta}$), which is the mass difference between the parent and daughter nuclei, including the electron mass.
Thus, selecting EDF that produces precise $Q_{\beta}$ is crucial for reliable $\beta$-decay predictions, in addition to correctly reproducing Gamow-Teller strength distributions.
However, choosing a specific functional from many available Skyrme functionals is not straightforward.
This raises the question: Is there any physical quantity like $G_{0}'$ that can serve as a clue to select a EDF that accurately reproduces the experimental values of $Q_{\beta}$.
The symmetry energy might provide an answer.
Wang et al. studied~\cite{WangNing2013} this relationship within the Skyrme Hartree-Fock framework and found a clear correlation between the proton-neutron Fermi energy difference ($\Delta\varepsilon=\varepsilon_{n}^{F}-\varepsilon_{p}^{F})$ and separation energy difference ($\Delta S=S_{p}-S_{n}$).
They also demonstrated that the average deviation $|\Delta \varepsilon-\Delta S|$ for $19$ doubly-magic and semi-magic nuclei exhibits a quadratic dependence on the symmetry energy.
A dependence of the symmetry energy on $Q_{\beta}$ was also found for odd-$A$ nuclei~\cite{Jianmin2013}.
While $Q_{\beta}$ is expected to correlate with the symmetry energy, this relationship has not yet to be fully clarified.
\par
The main purpose of this work is to study a correlation between the symmetry energy and $Q_{\beta}$ using Skyrme Hartree-Fock-Bogoliubov (Skyrme-HFB) calculations.
We filter functionals by comparing them with evaluated $Q_{\beta,\rm{AME}}$ data taken from AME2020~\cite{AME2020} and discuss whether the symmetry energy can be used as an indicator for selecting a functional suitable for systematical $Q_{\beta}$ calculations.
In parallel, we also study the relationships between $Q_{\beta}$ and other properties of Skyrme functionals, such as Landau parameters and nuclear bulk properties.
Sect.~\ref{sect:method} describes the method used in this work. 
Our results are presented in Sect.~\ref{sect:result}, and the conclusions are given in Sect.~\ref{sect:summary}.
%
%
%
\section{Method}
\label{sect:method}
We first show a relationship between $Q_{\beta}$, the Fermi energies, and the symmetry energy.
We often use the following definition of $Q_{\beta}$ for nuclei with neutron number $N$ and atomic number $Z$,
\begin{equation}
Q_{\beta}=\Delta M_{\mathrm{nH}}+B(N,Z)-B(N-1,Z+1).
\label{eq:Qbeta}
\end{equation}
Here, $B(N,Z)$ is the binding energy, and $\Delta M_{\mathrm{nH}}$ is mass difference between neutron and hydrogen atom.
Eq.~\eqref{eq:Qbeta} can be also expressed by 
\begin{equation}
Q_{\beta}=\Delta M_{\mathrm{nH}}+S_{p}(N,Z+1)-S_{n}(N,Z+1),
\label{eq:Qbeta2}
\end{equation}
where $S_{n}$ and $S_{p}$ are the neutron and proton separation energies, respectively.
From the liquid-drop mass formula, the difference between proton and neutron separation energies in Eq.~\eqref{eq:Qbeta2} can be expressed as~\cite{WangNing2013}
\begin{equation}
S_{p}(N,Z+1)-S_{n}(N,Z+1)
\simeq 2a_{c}\frac{Z}{A^{1/3}}-4a_{\rm{sym}}IA,
\label{eq:liquid}
\end{equation}
where $I=(N-Z)/A$.
Equations~\eqref{eq:Qbeta2} and \eqref{eq:liquid} indicate that $Q_{\beta}$ are dependent on the symmetry energy coefficient, $a_{\rm{sym}}$.
\par
When estimating $Q_{\beta}$ within EDF models, Equation~\eqref{eq:Qbeta} is used. 
At this stage, the binding energy $B(N,Z)$ of odd nuclei is essentially involved.
Although the calculation of $B(N,Z)$ of odd nuclei is practically possible within EDF models, it requires accounting for renormalization effects arising from the blocking effect and some treatments of the time-odd components of effective interactions~\cite{Bender2003, Kasuya2020}, which incur additional computational costs.
To avoid this issue, we often adopt a non-interacting quasiparticle approximation~\cite{Engel1999}, simplifying the system of odd nuclei by treating it as a core (even-even nuclei) plus valence quasiparticle(s).
In this scheme, $Q_{\beta}$ of even-even nuclei is approximated as~\cite{Engel1999}
\begin{equation}
Q_{\beta} \approx \Delta M_{\mathrm{nH}}-\lambda_{p}+\lambda_{n}-E_{2qp,lowest}.
\label{eq:Qbeta3}
\end{equation}
Here, the Fermi energies of the proton and neutron are represented by $\lambda_{p}$ and $\lambda_{n}$, respectively, and $E_{2qp,lowest}$ is the sum of the lowest quasiparticle energies of proton and neutron.
We will discuss the correlation between $Q_{\beta}$ given by Eq.~\eqref{eq:Qbeta3} and nuclear bulk properties including the symmetry energy.
\par
To calculate the quasiparticle and Fermi energies in Eq.~\eqref{eq:Qbeta3}, we employ the Skyrme-HFB model under the assumption of spherical symmetry.
The dependence of $Q_{\beta}$ on nuclear bulk properties is studied using the number of $N_f$ Skyrme functionals listed in Table~\ref{table:functionals}: $N_f=42$.
Most of these functionals are those recommended in Ref.~\cite{Bonasera2018}, covering a wide range of characteristics of Skryme forces.
In addition to these, we have added the SkP functional~\cite{SkP} and several modern functionals, namely SAMi~\cite{SAMi}, KIDS~\cite{KIDS0, KIDSfamily}, and UNEDF~\cite{UNEDF0, UNEDF1}.
For the pairing force in the Skyrme-HFB model, the Gogny-type finite-range interaction
\begin{equation}
V(\vec{r}_1,\vec{r}_2)
=f \sum_{i=1}^{2}
\Big(W_{i}+B_{i} P_{\sigma}-H_{i} P_{\tau}-M_{i} P_{\sigma} P_{\tau} \Big) e^{-r_{12}^{2}/\mu_{i}^{2}},
\label{eq:finiteforce}
\end{equation}
is used, where $r_{12} = |\vec{r}_{1} - \vec{r}_{2}|$.
The spin and isospin exchange operators are represented by $P_{\sigma}$ and $P_{\tau}$, respectively.
The parameters $W_{i}$, $B_{i}$, $H_{i}$, $M_{i}$, and $\mu_{i}$ are taken from the D1S force~\cite{Berger1984}.
The scaling factor $f$ is introduced in Eq.~\eqref{eq:finiteforce} to reduce uncertainties arising from the pairing correlation as explained later.
\par
The symmetry energy around the saturation density $\rho_{0}$ of these functionals can be expressed by
\begin{equation}
E_{\rm{sym}}(\rho)=J+Lx+\frac{1}{2}K_{\rm{sym}}x^{2}+\mathcal{O}(x^{3}),
\label{eq:Esym}
\end{equation}
where $\displaystyle x=\frac{\rho-\rho_{0}}{3\rho_{0}}$, $J=E_{\rm{sym}}(\rho_{0})$, $\displaystyle L=3\rho_{0}\left(\frac{\partial E_{\rm{sym}}}{\partial \rho}\right)_{\rho=\rho_{0}}$, and $\displaystyle K_{\rm{sym}}=9\rho_{0}^{2}\left(\frac{\partial^{2} E_{\rm{sym}}}{\partial \rho^{2}}\right)_{\rho=\rho_{0}}$.
The explicit forms of $E_{\rm{sym}}$, $L$, $K_{\rm{sym}}$ in terms of the Skyrme parameters, $t_{i}, x_{i} (i=0, 1, 2, 3)$, and $\alpha$ are given in the appendix.
\par
In addition to the symmetry energy, the relationship between the effective mass $m^{*}/m$ and the $Q_{\beta}$ value is also analyzed.
This is based on the fact that the effective mass governs the single-particle level density around the Fermi energy and significantly influences the two-quasiparticle energies in Eq.~\eqref{eq:Qbeta3}.
In general, functionals with a small $m^{*}/m$ yield sparse single-particle level densities around the Fermi energy, which are inconsistent with empirical data (see Sect.~2.7 of Ref.~\cite{RingandSchuck}), whereas functionals with $m^{*}/m\simeq1$ provide a reasonable single-particle level density.
As shown in Table~\ref{table:functionals}, the $42$ functionals exhibit a wide range of effective mass.
We categorize the functionals listed in Table~\ref{table:functionals} into two groups: those with a small effective mass $m^{*}/m \le 0.75$ ($N_{f,\le0.75}=12)$ and those with a large effective mass $m^{*}/m>0.75$ ($N_{f,>0.75}=30$).
\par
\begin{figure}
\includegraphics[width=0.99\linewidth]{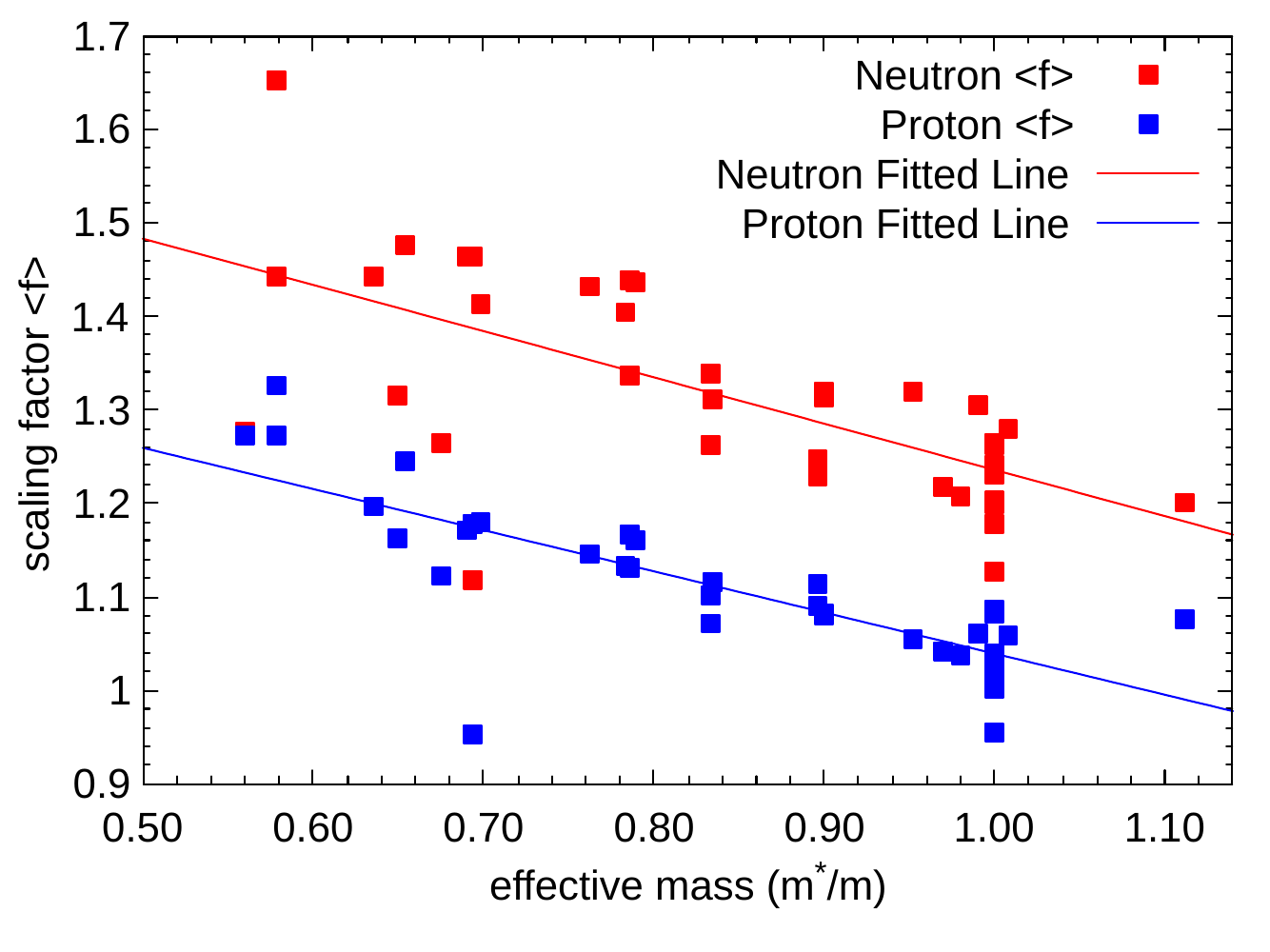}
\caption{Mean scaling factor $\langle f(x) \rangle $ for proton (blue) and neutron (red) pairing forces, where $x=m^{*}/m$.
The red and blue solid lines are the results of the least square fitting for neutron and proton $\langle f(x) \rangle$ (Eq.~\eqref{eq:scalingfactor}), respectively.}
\label{fig:pairinggap}
\end{figure}

We apply a special treatment to the pairing strengths.
If the pairing strengths in Eq.~\eqref{eq:finiteforce} are kept the same for every functional, the resulting pairing gaps will differ, affecting the two-quasiparticle energies in Eq.~\eqref{eq:Qbeta3}.
Since our goal is to study the correlation between $Q_{\beta}$, the symmetry energy, the effective mass, etc., we aim to minimize the variations in $Q_{\beta}$ arising from the pairing correlation.
To achieve this, we introduced a scaling factor $f=f(x)$, where $x \equiv m^{*}/m$, in Eq.~\eqref{eq:finiteforce}, which is determined to reproduce empirical pairing gaps.
We sampled $^{50-56}$Ca,$^{70-76}$Ni,$^{126-140}$Sn, $^{210-216}$Pb (19 nuclei) to calculate neutron pairing gaps and $^{66}$Fe, $^{80}$Zn, $^{82}$Ge, $^{84}$Se, $^{130}$Cd, $^{134}$Te,$^{204}$Pt, $^{206}$Hg (9 nuclei) for proton pairing gaps.
We determined the optimal $f$ that best reproduces the empirical pairing gaps estimated from the three-point mass difference~\cite{Satula1998} for each nucleus.
The results are shown in Fig.~\ref{fig:pairinggap}.
The red and blue squares represent the mean value of $f(x)$ defined by 
\begin{equation}
    \langle f(x) \rangle= \frac{1}{N_{\rm{paring}}} \sum_{\nu}^{N_{\rm{pairing}}} f(x,\nu),
\end{equation}
obtained from $N_{\rm{pairing}}=19$ for neutron gaps and $N_{\rm{pairing}}=9$ for proton gaps, respectively.
Here, $\nu$ indicate the nuclide mentioned above and $f(x,\nu)$ is the scaling factor that reproduces the empirical pairing gap for the nuclide $\nu$.
A clear linear dependence of $\langle f \rangle$ on the effective mass can be observed for both protons and neutrons.
This result is consistent with that reported in Ref.~\cite{Yamagami2009}.
The scaling factors are systematically greater than $1$.
In fact, relativistic-Hartree-Bogoliubov approaches conventionally use $f=1.15$~\cite{Gonzalez1996, Vretenar2005}.
As $m^{*}/m$ decreases, $f(x)$ increases because the single-particle level densities become sparse for small $x=m^{*}/m$, requiring a strong pairing force to reproduce the empirical pairing gaps.
We assume a linear dependence of $f(x)$ on effective mass and fit the parameters within a least squares fitting, resulting in
\begin{equation}
f(x)=
\begin{cases}
-0.4947x+1.7303 \quad {\rm for \: neutron}\\
-0.4401x+1.4792 \quad {\rm for \: proton}.
\end{cases}
\label{eq:scalingfactor}
\end{equation}
We present the result of our analysis performed using Eq.~\eqref{eq:scalingfactor} hereafter.
\begin{table*}
\centering
\caption{$42$ Functional used in this work and its properties for nuclear matter at the saturation density of $\rho_{0}$. The mean difference $\mu$ from the experimental $Q_{\beta}$ and the RMS deviation $\gamma$ are also shown.}
\begin{tabular}{l|r|r|r|r|r|r|r|r|r|r|r|r||r r||c}
\hline\hline
functional & \multicolumn{1}{c|}{$\rho_{0}$} & \multicolumn{1}{c|}{$E/A$} & \multicolumn{1}{c|}{$J$} & \multicolumn{1}{c|}{$L$} & \multicolumn{1}{c|}{$K_{\rm{sym}}$} & \multicolumn{1}{c|}{$K$} & $m^{*}/m$ & \multicolumn{1}{c|}{$W$} & \multicolumn{1}{c|}{$F_{0}$} & \multicolumn{1}{c|}{$F_{0}'$} & \multicolumn{1}{c|}{$G_{0}$} & \multicolumn{1}{c||}{$G_{0}'$} & \multicolumn{1}{c}{$\mu$} & \multicolumn{1}{c||}{$\gamma$} & Ref.\\
\hline
SkP & 0.1625 & $-15.95$ & 30.00 & $19.66$ & $-266.7$ & 201.0 & 1.000 & 150.0 & $-0.273$ & $0.559$ & $-0.229$ & $0.061$ & $0.361$ & 0.925 & \cite{SkP}\\
SkM* & 0.1602 & $-15.75$ & 30.04 & $45.80$ & $-155.9$ & 216.4 & 0.789 & 195.0 & $-0.229$ & $0.926$ & $0.325$ & $0.937$ & $-0.731$ & 1.247 & \cite{SkM*}\\
SGII & 0.1583 & $-15.58$ & 26.83 & $37.66$ & $-145.8$ &  214.4 & 0.786 & 157.5 & $-0.233$ & $0.728$ & $0.622$ & $0.934$ & $-1.848$ & 2.138 & \cite{Giai1981} \\
SLy4 & 0.1600 & $-15.93$ & 29.50 & $60.39$ & $-40.60$ & 247.7 & 0.650 & 185.8 & $-0.276$ & $0.815$ & $1.768$ & $0.881$ & $0.937$ & 1.425 & \cite{SLy}\\
SLy5 & 0.1603 & $-15.98$ & 32.03 & $48.27$ & $-112.3$ &  229.9 & 0.697 & 189.0 & $-0.276$ & $0.815$ & $1.123$ & $-0.141$ & $0.732$ & 1.264 & \cite{SLy} \\
SLy6 & 0.1590 & $-15.92$ & 31.96 & $47.45$ & $-112.7$ & 229.8 & 0.690 & 183.0 & $-0.280$ & $0.803$ & $1.408$ & $0.899$ & $1.009$ & 1.423 & \cite{SLy} \\
SkMP & 0.1570 & $-15.54$ & 29.88 & $70.30$ & $-49.82$ & 230.7 & 0.654 & 240.0 & $-0.310$ & $0.610$ & $0.797$ & $0.879$ & $-1.863$ & 2.321 & \cite{SkMP} \\
LNS & 0.1650 & $-15.28$ & 32.28 & $60.34$ & $-119.0$ & 194.8 & 0.834 & 144.0 & $-0.362$ & $1.147$ &$0.342$ & $0.939$ & $-0.031$ & 0.930 & \cite{LNS} \\
SkT1 & 0.1610 & $-15.98$ & 32.02 & $56.18$ & $-134.8$ & 236.1 & 1.000 & 165.0 & $0.064$ & $1.596$ & $-0.400$ & $0.164$ & $0.091$ & 0.880 & \cite{SkT}\\
SkT1* & 0.1602 & $-15.98$ & 32.02 & $56.11$ & $-135.2$ & 236.0 & 1.000 & 142.5 & $0.067$ & $1.605$ & $-0.396$ & $0.176$ & $-0.060$ & 0.767 & \cite{SkT}\\
SkT2 & 0.1610 & $-15.94$ & 32.00 & $56.16$ & $-134.7$ &  235.7 & 1.000 & 180.0 & $0.062$ & $1.594$ & $-0.406$ & $0.157$ & $0.177$ & 0.799 & \cite{SkT}\\
SkT3 & 0.1610 & $-15.94$ & 31.50 & $55.32$ & $-132.1$ & 235.7 & 1.000 & 189.0 & $0.062$ & $1.554$ & $-1.230$ & $0.452$ & $0.172$ & 0.786 & \cite{SkT}\\
SkT3* & 0.1602 & $-15.98$ & 31.68 & $55.85$ & $-132.2$ & 236.0 & 1.000 & 142.5 & $0.067$ & $1.578$ & $-1.226$ & $0.462$ & $-0.026$ & 1.000 & \cite{SkT}\\
SkT8 & 0.1607 & $-15.94$ & 29.92 & $33.73$ & $-187.5$ & 235.7 & 0.833 & 163.5 & $-0.115$ & $1.023$ & $0.007$ & $0.239$ & $0.280$ & 1.005 & \cite{SkT}\\
SkT9 & 0.1603 & $-15.88$ & 29.76 & $33.74$ & $-185.6$ & 234.9 & 0.833 & 195.0 & $-0.116$ & $1.016$ & $-0.010$ & $0.214$ & $0.248$ & 0.951 & \cite{SkT}\\
SkI3 & 0.1577 & $-15.96$ & 34.83 & $100.5$ & $72.92$ & 258.0 & 0.578 & 188.5 & $-0.320$ & $0.653$ & $1.894$ & $0.853$ & $0.350$ & 1.893 & \cite{SkI}\\
SkI4 & 0.1600 & $-15.93$ & 29.50 & $60.39$ & $-40.60$ & 247.7 & 0.650 & 185.8 & $-0.273$ & $0.559$ & $1.768$ & $0.881$ & $0.764$ & 1.667 & \cite{SkI}\\
SkI5 & 0.1557 & $-15.83$ & 36.63 & $129.3$ & $159.4$ & 255.6 & 0.579 & 185.4 & $-0.319$ & $0.757$ & $1.788$ & $0.853$ & $-0.731$ & 1.526 & \cite{SkI}\\
SV-bas & 0.1596 & $-15.90$ & 30.00 & $32.37$ & $-221.7$ & 233.4 & 0.900 &  158.7 & $-0.049$ & $1.201$ & $0.001$ & $0.989$ & $0.647$ & 1.162 & \cite{SV-min} \\
SV-min& 0.1611 & $-15.91$ & 30.66 & $44.81$ & $-156.6$ & 221.8 & 0.952 & 157.2 & $-0.050$ & $1.365$ & $0.581$ & $1.009$ & $0.613$ & 1.162 & \cite{SV-min} \\
SV-sym32 & 0.1595 & $-15.94$ & 32.00 & $57.07$ & $-148.8$ & 233.8 & 0.900 & 159.4 & $-0.046$ & $1.349$ & $-0.146$ & $0.990$ & $0.368$ & 1.023 & \cite{SV-min}\\
SV-m56-O & 0.1575 & $-15.81$ & 27.01 & $49.96$ &  $-45.04$ & 254.6 & 0.556 & 186.2 & $-0.352$ & $0.236$ & $1.777$ & $0.841$ & $0.380$ & 1.459 & \cite{SV-m}\\
SV-m64-O & 0.1588 & $-15.82$ & 27.01 & $30.63$ & $-144.8$ & 241.4 & 0.635 & 176.9 & $-0.303$ & $0.404$ & $1.297$ & $0.874$ & $0.558$ & 1.344 & \cite{SV-m}\\
NRAPR& 0.1606 & $-15.85$ & 32.78 & $59.64$ & $-123.3$ & 225.6 & 0.694 & 62.94 & $-0.294$ & $0.848$ & $-0.035$ & $0.410$ & $-0.240$ & 2.045 & \cite{NRAPR}\\
Z$_{\sigma}$ & 0.1627 & $-15.84$ & 26.72 & $-29.10$ & $-400.7$ & 232.9 & 0.783 & 185.5 & $-0.185$ & $0.682$ & $0.108$ & $0.395$ & $1.447$ & 1.877 & \cite{Zsig} \\
SAMi & 0.1586 & $-15.91$ & 28.17 & 43.70 & $-119.9$ & 244.8 & 0.675 & 158.0 & $-0.248$ & $0.557$ & $0.148$ & $0.350$ & $-1.311$ & 1.870 & \cite{SAMi}\\
SkO & 0.1604 & $-15.84$ & 31.97 & 79.15 & $-43.16$ & 223.3 & 0.896 & 154.4 & $-0.097$ & $1.328$ & $0.484$ & $0.984$ & $-0.077$ & 1.107 & \cite{SkO}\\
SkO' & 0.1602 & $-15.75$ & 31.95 & 68.93 &  $-78.81$ & 222.3 & 0.896 & 204.9 & $-0.099$ & $1.329$ & $-1.612$ & $0.792$ & $0.147$ & 1.106 & \cite{SkO} \\
KIDS0 & 0.1599 & $-15.98$ & 32.38 & 47.24 & $-160.4$ & 239.8 & 0.991 & 162.5 & $-0.233$ & $1.610$ & $-0.177$ & $1.030$ & $1.090$ & 1.411 & \cite{KIDS0}\\
KIDSA & 0.1599 & $-15.98$ & 32.59 & 63.98 & $-143.6$ & 229.8 & 1.005 & 148.0 & $-0.360$ & $1.664$ & $-0.996$ & $0.335$ & $-1.482$ & 1.773 & \cite{KIDSfamily} \\
KIDSB & 0.1599 & $-15.98$ & 31.60 & 56.01 & $-166.0$ & 239.8 & 0.997 & 148.4 & $-0.227$ & $1.564$ & $-0.862$ & $0.351$ & $-1.394$ & 1.648 & \cite{KIDSfamily} \\
KIDSC & 0.1599 & $-15.98$ & 30.59 & 55.98 & $-95.59$ & 249.8 & 0.983 & 152.0 & $-0.104$ & $1.447$ & $-0.716$ & $0.366$ & $-1.001$ & 1.299 & \cite{KIDSfamily} \\
KIDSD & 0.1599 & $-15.98$ & 29.59 & 44.98 & $-138.6$ & 259.8 & 0.970 & 152.5 & $0.017$ & $1.336$ & $-0.576$ & $0.380$ & $-0.874$ & 1.197 & \cite{KIDSfamily} \\
UNEDF0 & 0.1605 & $-16.06$ & 30.54 & 45.08 & $-189.7$ & 230.0 & 1.111 & 159.0 & $0.153$ & $1.757$ & $-0.757$ & $1.083$ & $0.322$ & 1.041 & \cite{UNEDF0} \\
UNEDF1 & 0.1587 & $-15.80$ & 28.99 & 40.01 & $-179.5$ & 220.0 & 1.008 & 148.1 & $0.008$ & $1.391$ & $-0.100$ & $1.034$ & $1.178$ & 1.523 & \cite{UNEDF1} \\
SQMC650 & 0.1650 & $-15.55$ & 32.93 & 53.07 & $-163.9$ & 205.6 & 0.786 & 187.3 & $-0.344$ & $1.063$ & $0.097$ & $0.925$ & $0.253$ & 0.913 & \cite{SQMC}\\
SQMC700 & 0.1650 & $-15.48$ & 32.77 & 58.59 & $-134.7$ & 209.6 & 0.762 & 177.4 & $-0.399$ & $0.991$ & $0.305$ & $0.913$ & $0.377$ & 1.111 & \cite{SQMC}\\
KDE0 & 0.1608 & $-16.10$ & 32.98 & 45.21 & $-144.8$ & 228.7 & 0.717 & 
 $193.4$ & $-0.261$ & 0.919 & 1.408 & 0.899 & 1.401 & 1.744 & \cite{KDE} \\
KDE0v1 & 0.1646 & $-16.23$ & 34.58 & 54.69 & $-127.1$ & 227.5 & 0.744 & $186.6$ & $-0.248$ & 1.057 & 0.669 & 0.001 & 1.492 & 1.754 & \cite{KDE} \\
SK255 & 0.1573 & $-16.33$ & 37.39 & 95.05 & $-58.32$ & 254.9 & 0.797 & $143.1$ & $-0.070$ & 1.455 & $-0.709$ & 0.373 & 0.227 & 1.194 & \cite{SK255} \\
MSL0 & 0.1600 & $-16.00$ & 30.00 & 60.00 & $-99.32$ & 230.0 & 0.800 & $200.0$ & $-0.168$ & 0.954 & $-0.524$ & 0.416 & $-1.901$ & 2.142 & \cite{MSL0} \\
Skxs20 & 0.1616 & $-15.80$ & 35.49 & 67.08 & $-122.3$ & 201.8 & 0.964 & $162.7$ & $-0.127$ & 1.765 & $-0.401$ & 0.129 & 1.399 & 1.858 & \cite{Skxs} \\
\hline\hline
\end{tabular}
\label{table:functionals}
\end{table*}
\section{Result}
\label{sect:result}
\subsection{Pearson linear correlation coefficients}
We study the correlation between nuclear bulk properties and the reproduction accuracy of $Q_{\beta}$ for the Skyrme functionals.
To this end, we apply the Pearson linear coefficient, which provides fundamental information on the relationship between two arbitrary quantities.
The Pearson linear correlation coefficients are defined as
\begin{equation}
    P(A,B)=\frac{\mathrm{C}(A,B)}{\sigma(A)\sigma(B)}.
    \label{eq:PearsonCoefficient}
\end{equation}
Here, $\mathrm{C}(A,B)$ is the covariance between quantity $A$ and $B$ given by
\begin{equation}
\mathrm{C}(A,B)=\frac{1}{N_{f}}\sum_{i}^{N_{f}}(A_{i}-\langle A \rangle)(B_{i}-\langle B \rangle)
\label{eq:covariance}
\end{equation}
and $\sigma(A)$ is the standard deviation of quantity $A$ represented by
\begin{equation}
\sigma(A)=\sqrt{\mathrm{C}(A,A)}.
\label{eq:standarddeviation2}
\end{equation}
The mean value of quantity $A$ is denoted by
\begin{equation}
    \langle A \rangle \equiv \frac{1}{N_{f}} \sum_{i}^{N_{f}} A_{i},
\end{equation}
where $A_{i}$ represents the corresponding value for functional $i$. 
The quantities $A$ and $B$ can represent various nuclear bulk properties, such as the saturation density $\rho_{0}$, the energy per nucleons $E/A$, the compressibility $K$, the symmetry energy $E_{\rm{sym}}$, the slope parameter of symmetry energy $L$, the quadratic parameter of symmetry energy $K_{\rm{sym}}$, the effective mass $m^{*}/m$.
If the Pearson linear correlation coefficients in Eq.~\eqref{eq:PearsonCoefficient} is close to $-1$, two quantities have a strong inverse proportionality.
On the other hand, if it is close to $+1$, two quantities have a strong proportionality.
\par
For the following discussion, we define $\gamma$ as the root-mean-square (RMS) deviation between the theoretically calculated and evaluated data $Q_{\beta,\rm{AME}}$ values:
\begin{equation}
\gamma=\sqrt{\frac{1}{N_{\rm{AME}}}\sum_{\nu}^{N_{\rm{AME}}} \left|\Delta Q_{\beta}^{(\nu)} \right|^{2}}.
\label{eq:rms}
\end{equation}
Here $\Delta Q_{\beta}^{(\nu)}=Q_{\beta,\rm{AME}}^{(\nu)}-Q_{\beta,\rm{calc}}^{(\nu)}$ with $Q_{\beta,\rm{calc}}^{(\nu)}$ being the calculated $Q_{\beta}$ of the nuclide $\nu$. 
One can regard $\gamma$ as a reproduction accuracy of $Q_{\beta,\rm{AME}}$ of Skyrme functionals.
We also define the mean deviation and the standard deviation as
\begin{equation}
\mu=\frac{1}{N_{\rm{AME}}}\sum_{\nu}^{N_{\rm{AME}}} \Delta Q_{\beta}^{(\nu)}
\label{eq:muQ}
\end{equation}
and
\begin{equation}
\sigma(\Delta Q_{\beta})=\sqrt{\frac{1}{N_{\rm{AME}}}\sum_{\nu}^{N_{\rm{AME}}} 
\left(\Delta Q_{\beta}^{(\nu)}-\mu\right)^{2}},
\label{eq:standarddeviation}
\end{equation}
respectively, which are also beneficial to find the correlations between the nuclear bulk properties.
For $Q_{\beta,\rm{AME}}^{(\nu)}$, there are about $1500$ available experimental data~\cite{AME2020}, from which we chose even-even nuclei with $Q_{\beta,\rm{AME}}^{(\nu)} \ge 0.5$ MeV from $^{20}$O to $^{246}$Pu, amounting $N_{\rm{AME}}=215$.
\par
\begin{figure*}
    \includegraphics[width=0.99\linewidth]{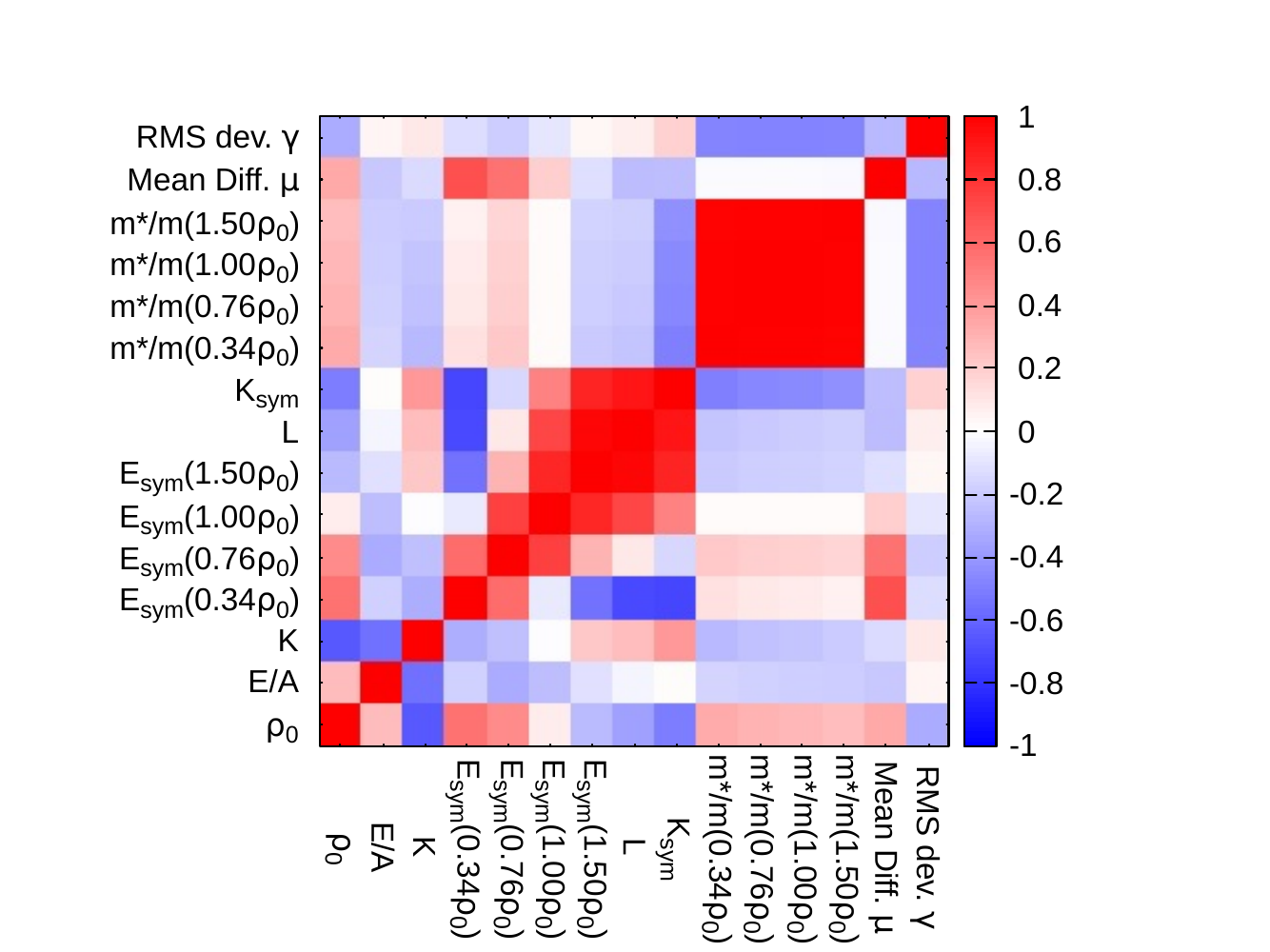}
    \caption{Pearson linear correlation coefficients between the mean difference $\mu$, the RMS deviation $\gamma$, and various nuclear properties. For effective mass $m^{*}/m$ and symmetry energy $E_{\rm{sym}}$, variations at various densities $\rho=0.34, 0.76, 1.00, 1.50\rho_{0}$ are also studied.}
    \label{fig:Pearson}
\end{figure*}
We first discuss the mean deviation $\mu$ and the RMS deviation $\gamma$ that are shown in Table~\ref{table:functionals}.
The SkT series and LNS exhibit relatively low $\gamma$ values ($<1.005$) and $\mu$ close to zero.
In particular, SkT1$^{*}$ provides the lowest $\gamma$ among the $42$ functionals.
SkT3$^{*}$ provides the most smallest $|\mu|$, but the RMS value $\gamma$ is larger than SkT1$^{*}$.
SkO and SkO', which are widely used for $\beta$-decay calculations, show $\mu$ closer to zero and $\gamma\sim1.100$.
These results are better than those of many other functionals, although their $\gamma$ values remain higher than those of SkT1-3 and LNS.
We focus on functionals that yield $\gamma>1.800$.
These include SGII, SkMP, SkI3, NRAPR, Z$_{\sigma}$, SAMi, MSL0, and Skxs20.
A common feature among these functionals is that either the symmetry energy coefficient $J$ or the effective mass $m^{*}/m$, or both are small.
We will discuss this point in more detail later.
\par
The results of the Pearson linear correlation coefficients are shown in Fig.~\ref{fig:Pearson}.
The red and blue sectors indicate positive and negative correlations, respectively.
We tested the correlation not only at the saturation density $\rho_{0}$, but also at $\rho=0.34\rho_{0}, 0.76\rho_{0}$ and $1.50\rho_{0}$.
The results show that $\gamma$ is largely insensitive to the symmetry energy at $\rho=0.34\rho_{0}, 0.76\rho_{0}, 1.00\rho_{0}$ and $1.50\rho_{0}$.
On the other hand, the mean difference $\mu$ shows a clear positive correlation with $E_{\rm{sym}}(\rho=0.34\rho_{0})$ and $E_{\rm{sym}}(\rho=0.76\rho_{0})$.
As density increases, the correlation between $\mu$ and $E_{\rm{sym}}(\rho)$ weakens and eventually becomes negative at $\rho=1.50\rho_{0}$. 
Additionally, $\gamma$ shows a negative correlation with effective masses at different densities, while $\mu$ does not exhibit any correlation with the effective masses.
The RMS deviation $\gamma$ also has a weak negative correlation with the saturation density, while $\mu$ exhibits a positive correlation.
\par
\begin{figure}
    \includegraphics[width=0.99\linewidth]{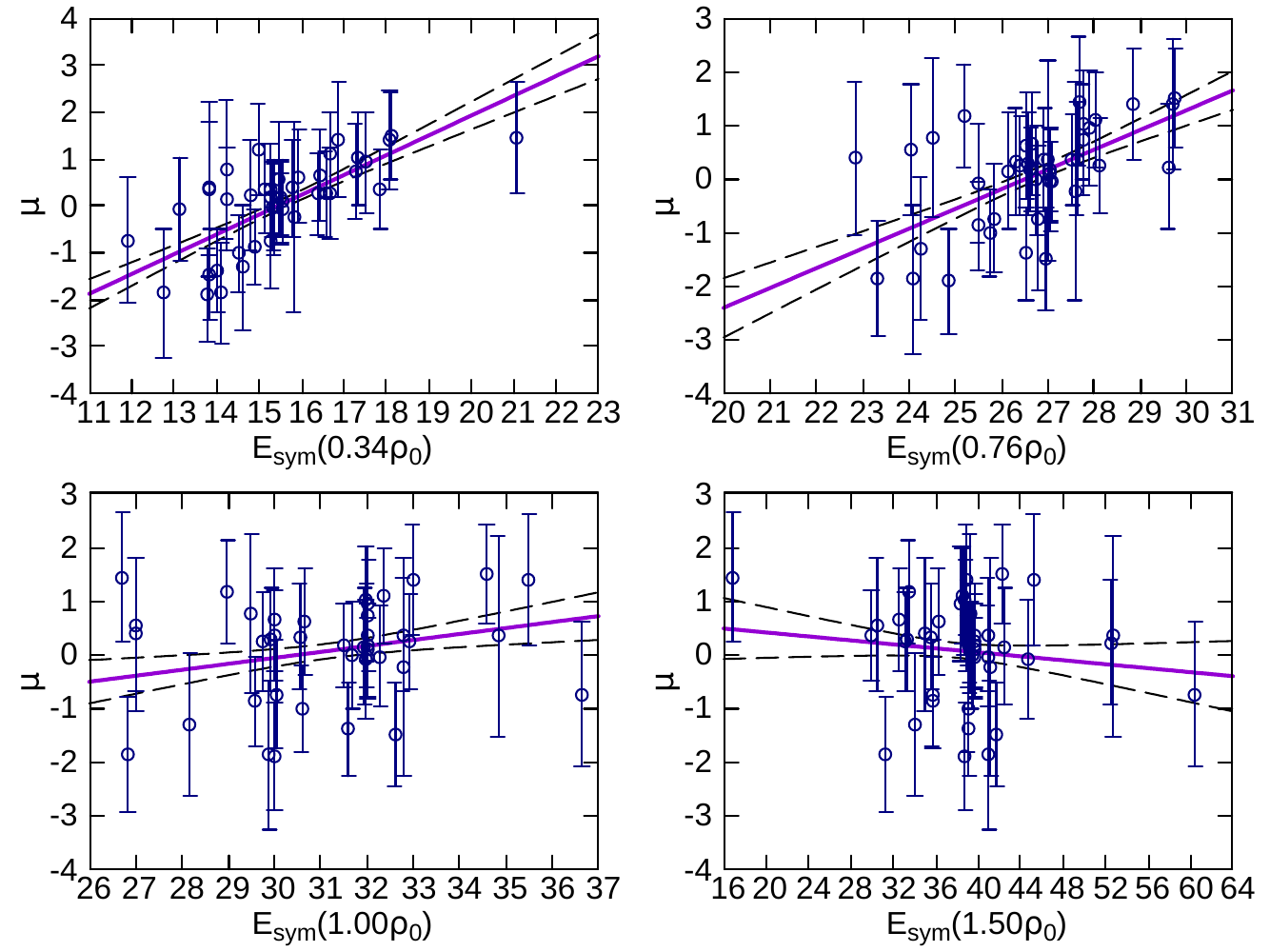}
    \caption{Mean difference ($\mu$) of experimental and calculated $Q_{\beta}$ as a function of symmetry energies at $\rho=0.34\rho_{0}, 0.76\rho_{0}, 1.00\rho_{0}$, and $1.50\rho_{0}$. The uncertainties for each circle are estimated by Eq.~\eqref{eq:standarddeviation}.
    The solid and dashed lines shows the mean value and the standard deviations estimated by the least-square fitting with a linear function, respectively.}
    \label{fig:MU-Esym}
\end{figure}
\begin{figure}
\includegraphics[width=0.99\linewidth]{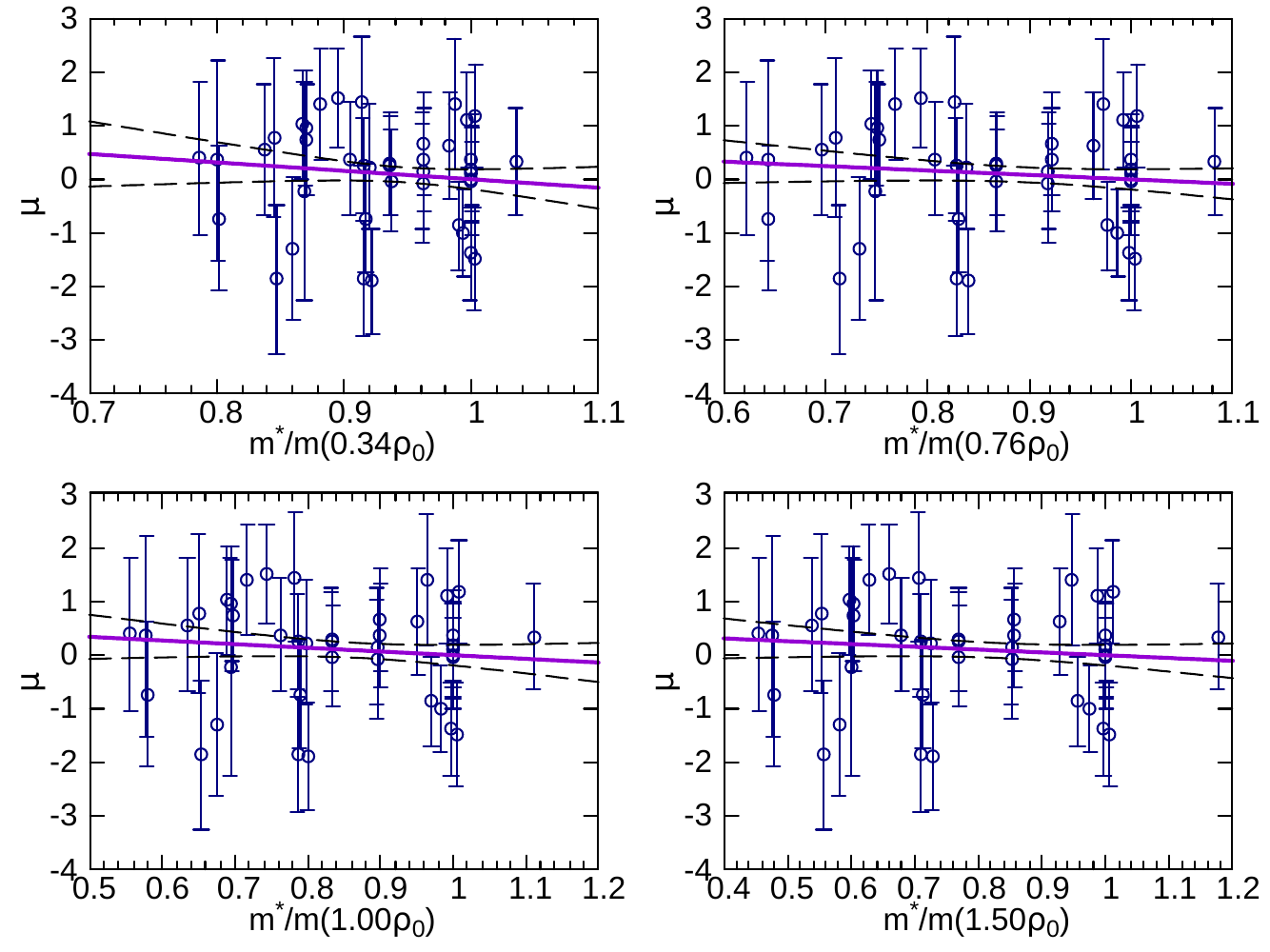}
\caption{Mean difference ($\mu$) of experimental and calculated $Q_{\beta}$ as a function of effective masses at $\rho=0.34\rho_{0}, 0.76\rho_{0}, 1.00\rho_{0}$, and $1.50\rho_{0}$.
The solid and dashed lines shows the mean value and the standard deviations estimated by the least-square fitting with a linear function, respectively.}
\label{fig:MU-M}
\end{figure}

To investigate in more detail the dependence of $\mu$ on the symmetry energy and the effective mass, we plot $\mu$ for different functionals as a function of symmetry energy and effective mass in Fig.~\ref{fig:MU-Esym} and Fig.~\ref{fig:MU-M}, respectively.
We first examine Fig.~\ref{fig:MU-Esym}.
As indicated by the Pearson linear correlation coefficients, a strong positive correlation is observed for $E_{\rm{sym}}(0.34\rho_{0})$.
The solid and dashed lines represent the mean value and the standard deviation estimated by a least-squares fit using a linear function.
This fit yields $\mu=0$ at $E_{\rm{sym}}(0.34\rho_{0})=15.47$~MeV.
A positive correlation remains at $\rho=0.76\rho_{0}$, giving $\mu=0$ at $E_{\rm{sym}}(0.76\rho_{0})=26.44$~MeV.
The correlation slope remains weakly positive at $\rho=1.00\rho_{0}$, but becomes negative at $\rho=1.50\rho_{0}$. 
We remind the reader of Eqs.~\eqref{eq:Qbeta2} and \eqref{eq:liquid}, which indicate that $Q_{\beta}$ increases with decreasing symmetry energy.
This trend is consistent with our findings.
Figure~\ref{fig:MU-M} shows the mean differences as functions of the effective mass.
As seen in Fig.~\ref{fig:Pearson}, $\mu$ does not exhibit any clear correlation with the effective mass.
\par
\begin{figure}
\includegraphics[width=0.99\linewidth]{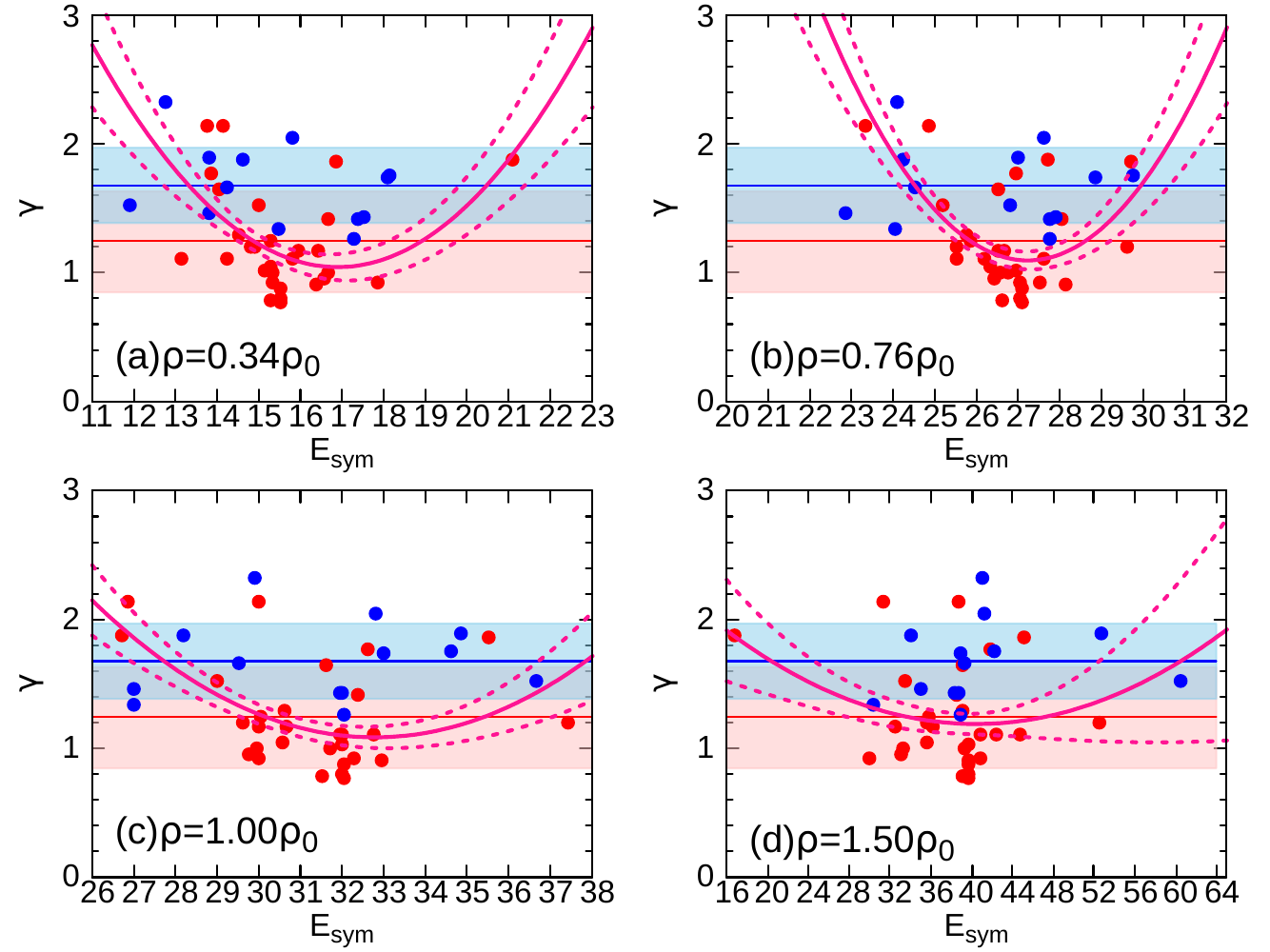}
\caption{RMS deviation ($\gamma$) for $42$ functionals as a function of symmetry energies at (a)$\rho=0.34\rho_{0}$, (b)$\rho=0.76\rho_{0}$, (c)$\rho=1.00\rho_{0}$, and (d)$\rho=1.50\rho_{0}$.
The red and blue circles indicate the functionals with $m^{*}/m < 0.75$ and $>0.75$, respectively.
The blue and red horizontal lines show $\langle \gamma \rangle=1.67$ and $1.24$ for functionals with $m^{*}/m \le 0.75$ and $>0.75$, respectively.
Similarly, the light-blue ($\gamma=1.67\pm0.29$) and light-red bands ($\gamma=1.24\pm0.39$) are the standard deviations, respectively.
The solid and dashed red quadratic curves show the mean values and the standard deviations estimated by the least-square fitting, respectively.}
\label{fig:RMS-Esym}
\end{figure}

We investigated the dependence of RMS deviation $\gamma$ on the symmetry energies.
The results are shown in Fig.~\ref{fig:RMS-Esym}.
We distinguish functionals based on their effective mass, as its correlation with the RMS value is negative, as found in Fig.~\ref{fig:Pearson}. 
The blue circles represent functionals with the effective mass $m^{*}/m \le 0.75$, while the red circles correspond to $m^{*}/m > 0.75$.
The Pearson linear correlation coefficients suggest that $\gamma$ has only a weak correlation with the symmetry energies.
However, we observe that functionals with $m^{*}/m>0.75$ tend to yield relatively small $\gamma$, whereas those with $m^{*}/m\le0.75$ generally result in larger $\gamma$.
The blue and red horizontal lines in Fig.~\ref{fig:RMS-Esym} show $\langle \gamma \rangle$ for functionals with $m^{*}/m \le 0.75$ and $>0.75$, respectively.
We notice a significant difference in their distribution even when considering the standard deviations, shown by the light-blue ($\gamma=1.67\pm0.29$) and light-red bands ($\gamma=1.24\pm0.39$).
\subsection{Mean symmetry energy}
At different densities, smaller values of $\gamma$ tend to cluster around specific values of $E_{\rm{sym}}$, such as $E_{\rm{sym}}\simeq 15.5$ and $32.5$~MeV at $\rho=0.34\rho_{0}$ and $1.00\rho_{0}$, respectively.
Moving away from these regions, both for lower and higher $E_{\rm{sym}}$, $\gamma$ tends to increase.
This outcome seems to contradict the Pearson linear correlation coefficient result, which indicated that $\gamma$ does not have a strong correlation with $E_{\rm{sym}}$ (Fig.~\ref{fig:Pearson}).
However, the key reason for this apparent contradiction is that $\gamma$ and $E_{\rm{sym}}$ exhibit a quadratic correlation rather than a linear correlation.
This result is consistent to what is reported in Ref.~\cite{WangNing2013} although the number of nuclei studied is much smaller than this study. 
To quantify this trend, we approximate the distribution of $\gamma$ for functionals with $m^{*}/m>0.75$ using a quadratic curve and determine the best fit via the least-square fitting.
The results are shown by the solid red quadratic curves in Fig.~\ref{fig:RMS-Esym} with the standard deviation indicated by the dashed curves.
From the fitting, we obtain the minimum RMS value at $E_{\rm{sym}}=16.9\pm0.4$~MeV at $\rho=0.34\rho_{0}$, $E_{\rm{sym}}=27.2\pm0.3$~MeV at $\rho=0.76\rho_{0}$, $E_{\rm{sym}}=32.8\pm0.7$~MeV at $\rho=1.00\rho_{0}$, and $E_{\rm{sym}}=40.4\pm6.5$~MeV at $\rho=1.50\rho_{0}$. 
We performed the same analysis for the slope parameter $L$, but could not determine the parameters of quadratic function as uniquely as we did for $E_{\rm{sym}}$.
Consequently, the mean difference and RMS deviation are primarily sensitive to $E_{\rm{sym}}$ among the selected nuclear matter properties.
\par
\begin{figure}
\includegraphics[width=0.99\linewidth]{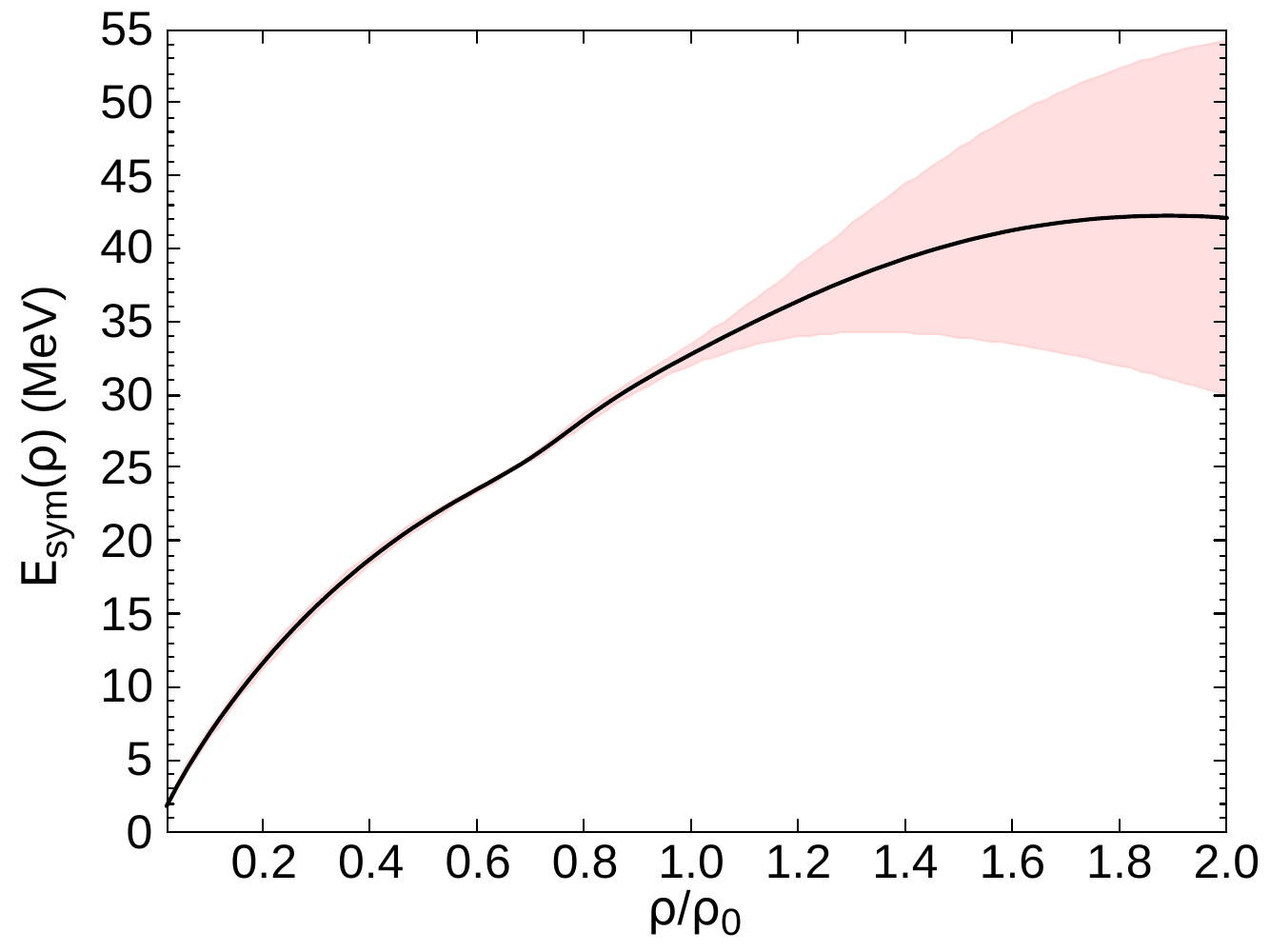}
\caption{Optimal symmetry energy $E_{\rm{sym}}$, determined by the fit of the quadratic curve, as a function of normalized density $\rho/\rho_{0}$. The standard deviations are shown by the shaded bands.}
\label{fig:Esymrho2}
\end{figure}

The analysis using the fit of the quadratic curve suggests an optimal $E_{\rm{sym}}$ that minimizes $\gamma$ for each density.
Figure~\ref{fig:Esymrho2} presents the optimal symmetry energy $E_{\rm{sym}}$ as a function of the normalized density $\rho/\rho_{0}$, with shaded bands indicating the standard deviations.
At low densities $(\rho<1.00\rho_{0})$, the standard deviations are relatively small, meaning that $E_{\rm{sym}}$ can be determined with high precision.
However, as the density increases, the standard deviations become larger, making it harder to pinpoint an exact value for $E_{\rm{sym}}$.
This trend is directly related to the curvature of the fitted quadratic curves.
As seen in Fig.~\ref{fig:RMS-Esym}, the quadratic curves have a steep curvature at $\rho=0.34\rho_{0}$, but a flatter curvature at $\rho=1.50\rho_{0}$.
A steeper curvature enables us to limit the range of optimal $E_{\rm{sym}}(\rho)$.
In contrast, a flatter curvature at high densities suggests that $\gamma$ is relatively insensitive to changes in $E_{\rm{sym}}$, making it difficult to constrain $E_{\rm{sym}}$ in this region.  
This result indicates that experimental $Q_{\beta}$ values can be used to constrain $E_{\rm{sym}}(\rho)$ at low densities, provided functionals with relatively high effective masses are chosen.  
However, at high densities, this approach is less effective, as $E_{\rm{sym}}$ does not exhibit clear linear or quadratic correlations with the RMS deviation $\gamma$.
\subsection{Reproduction accuracy for chosen functionals}
\label{sect:Reproduction}
\begin{figure}
\includegraphics[width=0.99\linewidth]{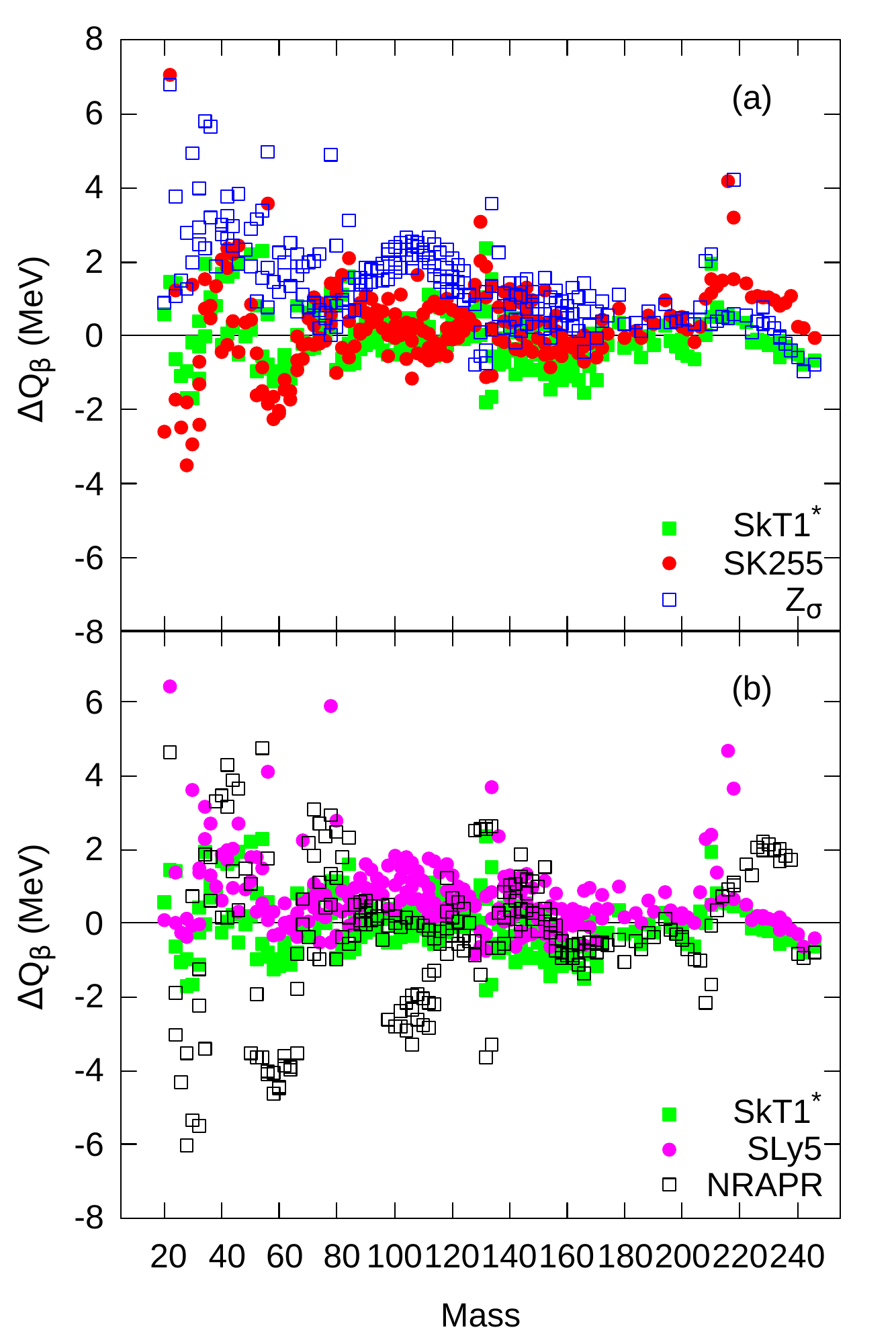}
\caption{$\Delta Q_{\beta}^{(\nu)}=Q_{\beta,\rm{AME}}^{(\nu)}-Q_{\beta,\rm{calc}}^{(\nu)}$ as a function of mass, comparing with (a)SkT1$^{*}$, SK255, and SGII, and (b)SkT1$^{*}$, SLy5, and NRAPR.}
\label{fig:Distribution}
\end{figure}

In the previous section, we identified that the optimal $E_{\rm{sym}}$ minimizing $\gamma$ corresponds to functionals with $32.8\pm0.7$~MeV in the effective mass range $m^{*}/m>0.75$.
Among the $42$ functionals examined, LNS, SkT family, KIDS0, KIDSA, SQMC650, SQMC700 closely match these conditions.
Except for KIDS0 and KIDSA, these functionals yield a mean deviation $|\mu|<0.4$~MeV and a relatively small RMS deviation $\gamma$.
To illustrate this further, Fig.~\ref{fig:Distribution}(a) shows the distribution of $\Delta Q_{\beta}^{(\nu)}$ for SkT1$^{*}$, SK255 and Z$_{\sigma}$. 
Here, SkT1$^{*}$ is selected from SkT family, while SK255 and Z$_{\sigma}$ represent extreme cases, possessing the largest and smallest $E_{\rm{sym}}$ values among the $42$ functionals, respectively.
The results for SkT1$^{*}$ show that $\Delta Q_{\beta,\rm{calc}}^{(\nu)}$ is distributed close to zero, indicating reasonable agreement with the experimental data.
SK255 exhibits a trend similar to SkT1$^{*}$ for many nuclei; however, it underestimates experimental data in the heavy mass region $A>200$.
Z$_{\sigma}$, on the other hand, shows a clear systematic underestimation of experimental data from light to $A\sim200$ nuclei.
As discussed previously in Fig.~\ref{fig:MU-Esym}, $\mu$ and the symmetry energy exhibit a positive correlation, indicating that smaller symmetry energies result in larger values of $Q_{\beta,\rm{calc}}^{(\nu)}$.
Therefore, Z$_{\sigma}$ globally underestimate the experimental data.
These findings further support the conclusion that SkT1$^{*}$ is a promising choice for accurate $Q_{\beta}$ predictions, reinforcing the importance of selecting functionals with appropriate symmetry energy and effective mass values.
\par
To investigate the influence of effective mass on $Q_{\beta}$, we also compare the result of SkT1$^{*}$ with those of two other functionals: SLy5 and NRAPR.
These functionals share a similar symmetry energy ($E_{\rm{sym}}=32.8\pm0.7$~MeV), but exhibit small effective masses ($m^{*}/m<0.75$).
The results obtained by SLy5 are close to those of SkT1$^{*}$, although some data points deviate toward higher values of $Q_{\beta}^{(\nu)}$ for certain nuclei.
This leads to a large mean difference $\mu$, as shown in Table~\ref{table:functionals}.
In contrast, NRAPR exhibits large fluctuations in $\Delta Q_{\beta}^{(\nu)}$ for specific mass regions around $A\sim30, 60, 110, 230$.
Although its mean difference is smaller than that of SLy5, it yields a large RMS deviation.
We note that NRAPR features an extremely weak spin--orbit interaction (see Table.~\ref{table:functionals}), which may affect the single-particle level structure and, consequently, the quasiparticle energies.
We also examined the results of other functionals with symmetry energies similar to that of SLy5 and relatively small effective mass, including SLy6, SQMC650, SQMC700, and KDE0.
Our analysis confirms that these functionals consistently produce deviations toward higher $Q_{\beta}^{(\nu)}$ values for nuclei similar to those observed with SLy5, particularly around magic and semi-magic nuclei.
Specific examples include $^{22}$O, $^{30,32}$Ne, $^{34,36}$Mg, $^{34,36,42}$Si, $^{44,46}$Ar, $^{56}$Ca, $^{68,78,80}$Ni, $^{132,134,136}$Sn, and $^{208,210,216,218}$Pb.

\section{Summary}
\label{sect:summary}
We studied the correlations between the symmetry energy and $Q_{\beta}$ using Skyrme-HFB calculations.
By comparing the functionals with the evaluated $Q_{\beta,\rm{AME}}$ data, we identified the optimal symmetry energy range suitable for systematic $Q_{\beta}$ calculations.
Furthermore, we examined the relationships between $Q_{\beta}$ and other nuclear bulk properties derived from $42$ Skyrme functionals using Pearson linear correlations.
\par
From the analysis of Pearson linear correlations, the RMS deviation is largely insensitive to the symmetry energy both at low and high densities.
The mean difference showed a positive correlation with $E_{\rm{sym}}(\rho=0.34\rho_{0})$ and $E_{\rm{sym}}(\rho=0.76\rho_{0})$, while the correlation weakened as the density increased. 
The RMS deviation showed a negative correlation with effective masses, while the mean difference does not exhibit any correlation with the effective masses.
Distinguishing functionals based on their effective mass, we observed that functionals with effective mass greater than $0.75$ tended to yield relatively small RMS deviations.
Using the least square fitting, we identified that the optimal $E_{\rm{sym}}$ minimizing the RMS deviation is $32.8\pm0.7$~MeV in the effective mass range $m^{*}/m>0.75$.
\par
We picked up SkT1$^{*}$ as a reference satisfying the optimal $E_{\rm{sym}}$ and the effective mass, and SK255 and Z$_{\sigma}$ represent extreme cases, possessing the largest and smallest $E_{\rm{sym}}$ among $42$ functionals, respectively.
The results for SkT1$^{*}$ show that $\Delta Q_{\beta,\rm{calc}}^{(\nu)}$ is distributed close to zero, indicating a reasonable agreement with the experimental data.
SK255 underestimates experimental data, particularly in the heavy mass region, and Z$_{\sigma}$ shows a systematic underestimation.
To investigate the influence of effective mass on $Q_{\beta}$, we also compared the result of SkT1$^{*}$ with two other functionals: SLy5 and NRAPR that share a similar symmetry energy to SkT1$^{*}$, but have small effective mass.
Deviation of $Q_{\beta}^{(\nu)}$ were observed particularly around around magic and semi-magic nuclei.
\par
For practical $\beta$-decay calculations, the $\beta$-strength function is important in addition to $Q_{\beta}$.
As mentioned in Sect.~\ref{sect:intro}, the Landau-Migdal parameter $G_{0}'$ has a significant impact on the $\beta$-strength function.
In this regard, LNS, SQMC650, and SQMC700 satisfy the condition for $E_{\rm{sym}}$ and also have relatively large $G_{0}'$ values.
On the other hand, the SkT family provides small $G_{0}'$ values, suggesting that these functionals may not be ideal candidates, despite their ability to systematically reproduce $Q_{\beta}$ reasonably well.
To properly study the $\beta$-strength function, a beyond-mean-field approach, such as the random-phase-approximation (RPA), is required.
Therefore, we do not discuss this topic in detail in the present paper.
In addition, our theoretical framework assumes spherical nuclear shapes.
However, many nuclei are known to be deformed, and it is likely that some of the $215$ nuclei considered in this study also exhibit deformation. 
If nuclear deformation is taken into account, the present results may differ to some extent.
Further studies, including analyses of deformation effects and the $\beta$-strength function, are currently in progress.


\begin{acknowledgments} 
The authors thank H. Nakada (Chiba University) for valuable comments.
This work was supported by JSPS KAKENHI Grant Numbers JP19K03824, JP23K03426, JP23H05434,  JP24K00647, and JP25H01269, 
the JSPS/NRF/NSFC A3 Foresight Program ``Nuclear Physics in the 21st Century'', and 
JST ERATO Grant No. JPMJER2304.
\end{acknowledgments}
\appendix
\section*{Appendix}
\label{appendix}
In Skyrme energy density functionals, the symmetry energy as a function of nuclear density is defined as
\begin{equation}
\begin{split}
E_{\rm{sym}}(\rho)
&=\frac{1}{3}\frac{\hbar^{2}}{2m}b\rho^{2/3}-\frac{t_{0}}{8}(1+2x_{0})\rho\\
&-\frac{\Theta_{\rm{sym}}}{24}b\rho^{5/3}
-\frac{t_{3}}{48}(1+2x_{3})\rho^{\alpha+1},
\end{split}
\end{equation}
where $\Theta_{\rm{sym}}=3t_{1}x_{1}-t_{2}(4+5x_{2})$ and $b=(3\pi^{2}/2)^{2/3}$.
The slope parameter at the saturation density is
\begin{equation}
\begin{split}
L&=\frac{\hbar^{2}}{3m}b\rho_{0}^{2/3}
-\frac{3t_{0}}{8}(2x_{0}+1)\rho_{0}
-\frac{5\Theta_{\rm{sym}}}{24}b\rho_{0}^{5/3}\\
&-\frac{t_{3}}{16}(2x_{3}+1)(\alpha+1)\rho_{0}^{\alpha+1}.
\end{split}
\end{equation}
The comprehensibility parameter of symmetry energy at the saturation density is
\begin{equation}
\begin{split}
K_{\rm{sym}}
&=-\frac{\hbar^{2}}{3m}b\rho_{0}^{2/3}
-\frac{5\Theta_{\rm{sym}}}{12}b\rho_{0}^{5/3}\\
&-\frac{3t_{3}}{16}(2x_{3}+1)\alpha(\alpha+1)\rho_{0}^{\alpha+1}.
\end{split}
\end{equation}
The effective mass is defined as
\begin{equation}
\frac{m}{m^{*}}=1+\frac{m}{8\hbar^{2}}
\Big(3t_{1}+t_{2}(5+4x_{2})\Big)\rho.
\end{equation}
\bibliography{ref}

\end{document}